\documentclass[conference]{IEEEtran}

\newtheorem{theo}{Theorem}

\newtheorem{lemm}{Lemma}
\newtheorem{exam}{Example}
\newtheorem{defi}{Definition}
\usepackage{epsfig}
\usepackage{amsmath}
\usepackage{amssymb}
\usepackage{subfigure}
\usepackage{graphicx}
\usepackage{epstopdf}
\usepackage{balance}
\newcommand \col {\rm col}
\newcommand{\diag}{\rm diag}

\ifCLASSINFOpdf

\else

\fi

\begin{document}
\title{\Large Event-based Bipartite Consensus on Signed Networks}
\author{\IEEEauthorblockN{Yulong Zhou}
\IEEEauthorblockA{School of Automation Engineering\\
University of Electronic\\
 Science and Technology of China\\
Chengdu 611731, China}
\and
\IEEEauthorblockN{Jiangping Hu}
\IEEEauthorblockA{School of Automation Engineering\\
University of Electronic \\
Science and Technology of China\\
Chengdu 611731, China}}

\maketitle

\begin{abstract}
In this paper, a bipartite consensus problem for a multi-agent system is formulated firstly. Then an event-based interaction rule is proposed for the multi-agent system with antagonistic interactions. The bipartite consensus stability is analyzed on the basis of spectral properties of the signed Laplacian matrix associated with multi-agent networks. Some simulation results are presented to illustrate the bipartite consensus with the proposed interaction rule.
\end{abstract}

\IEEEpeerreviewmaketitle

\section{Introduction}

The problem of reaching a consensus among a group of agents using only local interactions has a long history \cite{lyn,vic} and in recent years it has received a remarkable attention from different perspectives, such as engineering, computer science, biology, ecology and social sciences \cite{jad,olfa,reyn,couz,hegs}.

Till today, a large number of results have been obtained to solve consensus problem, whose common feature is the focus on cooperative system. In several real world cases, however, it is more reasonable to assume that both collaborative and adversary relationships exist in the multi-agent interactions. Some examples can be found in social network theory, where the multi-agent networks have antagonistic relationships \cite{wass,eas}. Very recently, a bipartite consensus problem is investigated with the help of signed network theory in \cite{alta}, where all agents converge to a value which is the same in modulus but different in sign. Though the interaction rule is Laplacian-like scheme, but a signed Laplacian matrix associated with the antagonistic network has some distinctive spectral properties.

In this paper, we consider a bipartite consensus for multi-agent systems with antagonistic networks by applying an event-based update strategy. Though some event-triggered consensus controls have been proposed for cooperative multi-agent systems \cite{dimos,hu,liu}, however, few results can be found for multi-agent systems with antagonistic interactions. In this paper, the bipartite consensus stability of the multi-agent system is analyzed for undirected and directed signed networks, respectively. Simultaneously, a positive lower bound is given for the inter-event times.

The rest of this paper is organized as follow. Section \ref{pre} presents some notations in signed graph theory and formulates a bipartite consensus problem. Then an event-based consensus strategy is proposed to solve the bipartite consensus problem in Section \ref{analysis}. Some numerical examples are given in Section \ref{simu} and a summary of the results is concluded in Section \ref{conc}.

\section{Preliminaries and Problem Formulation}\label{pre}

In this section, some basic concepts and notations in the signed graph theory will be firstly introduced. Then, an event-based consensus problem will be formulated for a multi-agent systems with antagonistic interactions.

\subsection{Signed Graph Theory}

A signed graph is a triple $\mathcal{G}=\{\mathcal{V},\mathcal{E},\mathcal{A}\}$, where $\mathcal{V}=\{v_1,v_2,\dots,v_n\}$ is a set of nodes with the indices belonging to a finite index set $\mathcal{N}=\{1,2,\dots,n\}$, $\mathcal{E}\subseteq \mathcal{V}\times \mathcal{V}$ is a set of edges, and $\mathcal{A}=[a_{ij}]\in \mathbb{R}^{n\times n}$ is an adjacency matrix of the weighted graph. If the edge $(i,j)\in \mathcal{E}$, then the element $a_ij$ of $\mathcal{A}$ is nonzero. $\mathcal{G}$ is called simple if it has no loops or multiple edges. A (directed) path is a sequence of edges in a (directed) graph of the form $(i_1,i_2), (i_2,i_3), \cdots, (i_{l-1},i_l)$ with distinct nodes. If there is a (directed) path between every two distinct nodes, $\mathcal{G}$ is said to be (strongly) connected. A semipath is defined as a sequence of nodes $i_1,\cdots, i_l$ such that $(i_\kappa, i_{\kappa-1})$ or $(i_{\kappa-1}, i_\kappa)$ belongs to the set $\mathcal{E}$. A (directed) cycle is a (directed) path beginning and ending with the same nodes $i_1=i_l$. A semicycle is a semipath with identical starting and ending nodes. For the signed graph $\mathcal{G}$, there exists a signal mapping $\sigma:\mathcal{E}\to\{+,-\}$ such that the edge set $\mathcal{E}=\mathcal{E}^+\cup\mathcal{E}^-$. The sets of positive and negative edges can be denoted by $\mathcal{E}^+=\{(i,j)|a_{ij}>0\}$ and $\mathcal{E}^-=\{(i,j)|a_{ij}<0\}$, respectively. In this paper,we assume that the weights between nodes $v_i$ and $v_j$ satisfy $a_{ij}a_{ji}\geq0$. The neighbor index set of node $v_i$ is denoted by $\mathcal{N}_i=\{j\in\mathcal{N}|(v_i,v_j)\in\mathcal{E}\}$. Define the row connectivity matrix and the column connectivity matrix $C_r=diag\{c_{r,1},c_{r,2},\dots,c_{r,n}\}, C_c=diag\{c_{c,1},c_{c,2},\dots,c_{c,n}\}$ where $c_{r,i}=\sum\limits_{j\in\mathcal{N}_i}|a_{ij}|$ and  $c_{c,i}=\sum\limits_{j\in\mathcal{N}_i}|a_{ji}|$, respectively. Obviously,$C_r=C_c$ for undirected graphs. For directed graphs, if $C_r=C_c$, the directed graph $\mathcal{G}$ is said to be weight balanced. The signed Laplacian matrix of a directed signed graph $\mathcal{G}$ is defined as
\begin{equation}\label{siglp}
L=C_r-\mathcal{A}
\end{equation}
which is an asymmetric matrix generally. For the undirected graph $\mathcal{G}$, the signed Laplacian matrix is defined analogously, however, is a symmetric matrix.
\begin{defi}
A signed graph $\mathcal{G}$ is said to be (strongly) connected if there is a (directed) path between every two distinct nodes.
\end{defi}
\begin{defi}
A signed graph $\mathcal{G}$ is said to be structurally balanced if it admits a bipartition of the nodes $\mathcal{V}_1$,$\mathcal{V}_2$,where $\mathcal{V}_1\cup\mathcal{V}_2=\mathcal{V}$ and $\mathcal{V}_1\cap\mathcal{V}_2=0$, such that $a_{ij}\geq 0$ for $\forall v_i, v_j\in\mathcal{V}_q \;(q\in\{1, 2\})$ and $a_{ij}\leq 0$ for $\forall v_i\in\mathcal{V}_q, v_j\in\mathcal{V}_r, q\neq r\; (q, r\in\{1,2\})$. Otherwise, it is said structurally unbalanced.
\end{defi}

\subsection{Problem Formulation}

Let us consider a group of agents whose dynamics are described by
\begin{equation}\label{dyna}
\dot{x}_i=u_i, i\in\mathcal{N},
\end{equation}
where $x_i\in \mathbf{R}$ is the state of agent $i$. Assume that the $n$ agents are belonging to two hostile groups $\mathcal{V}_1$ and $\mathcal{V}_2$, where $\mathcal{V}_1\cup\mathcal{V}_2=\mathcal{V}$ and $\mathcal{V}_1\cap\mathcal{V}_2=\O$. Thus, it is convenient to describe the interaction network as a structurally balanced signed graph $\mathcal{G}$.

In this paper, an event-based consensus problem will be considered for multi-agent systems with antagonistic interactions. To be specific, all agents update their states only at a series of event-times, which are determined by the measurement errors and the thresholds. Suppose that all agents have the same threshold and there is a sequence of event-times $t_0, t_1, \cdots,$ defined for them. Between two consecutive event-times, the interaction rule keeps a constant, i.e.,
\begin{equation*}
u_i(t)=u_i(t_l), \forall t\in [t_l,t_{l+1}), l=0, 1, \cdots.
\end{equation*}
In the networks with antagonistic links, an interaction law is proposed for agent $i$ by
\begin{equation}\label{cont}
u_i(t)=\sum\limits_{j\in\mathcal{N}_i}a_{ij}\big[x_j(t_l)-sgn(a_{ij})x_i(t_l)\big], l=0, 1, \cdots.
\end{equation}
where $sgn(\cdot)$ is a sign function.

\section{Event-based Antagonistic Consensus}\label{analysis}

Define a concatenation vector $x={\col}(x_1, \cdots, x_n)\in \mathbb{R}^{mn}$. Then the collective dynamics of the $n$ agents have a compact form by applying the interaction rule (\ref{cont})
\begin{equation}\label{comp}
\dot{x}(t)=-Lx(t_l), t\in[t_l, t_{l+1}), l=0, 1, \cdots,
\end{equation}
where $L$ is a signed Laplacian matrix defined by (\ref{siglp}). Define a measurement error $e(t)=x(t_l)-x(t)$ for $t\in[t_l, t_{l+1})$. Then the system (\ref{comp}) is changed to
\begin{equation}\label{compe}
\dot{x}(t)=-L\Big(x(t)+e(t)\Big), t\in[t_l, t_{l+1}), l=0, 1, \cdots.
\end{equation}

\subsection{Undirected Signed Networks}

\begin{lemm}\label{lem1}
If an undirected signed graph $\mathcal{G}$ is connected and structurally balanced, all eigenvalues of the signed Laplacian matrix $L$, defined by (\ref{siglp}), are nonnegative, i.e., $\lambda_1(L)=0, \lambda_2(L)>0$.
\end{lemm}

When the multi-agent network $\mathcal{G}$ is structurally balanced, all agents can be divided into two subsets $\mathcal{V}_1$ and $\mathcal{V}_2$. Define a diagonal matrix $D={\diag}\{d_1, \cdots, d_n\}$ such that $d_i=1$ if $i\in \mathcal{V}_1$ and $d_i=-1$ if $i\in \mathcal{V}_2$.

Let $\tilde{x}=Dx$ and $\tilde{e}=De$. Then the system (\ref{compe}) can be rewritten as follows
\begin{equation}\label{compe1}
\dot{\tilde{x}}(t)=-\tilde{L}\tilde{x}(t)-\tilde{L}\tilde{e}(t), t\in[t_l, t_{l+1}), l=0, 1, \cdots,
\end{equation}
where $\tilde{L}=DLD$.

According to Lemma \ref{lem1}, when the network is undirected, the signed Laplacian matrix $L$ is a symmetric positive semi-definite matrix. Thus, a candidate ISS Lyapunov function is chosen for the multi-agent system (\ref{compe1}):
\begin{equation}
V(\tilde{x})=\frac{1}{2}\tilde{x}^T\tilde{L}\tilde{x}.
\end{equation}

The derivative of $V(x)$ along the system (\ref{compe1}) is:
$$\begin{aligned}
\dot{V}&=-\tilde{x}^T\tilde{L}\tilde{L}\Big(\tilde{x}+\tilde{e}\Big)\\
 &=-\|\tilde{L}\tilde{x}\| ^2-\tilde{x}^T\tilde{L}\tilde{L}\tilde{e}\\
 &\leq -\|\tilde{L}\tilde{x}\| ^2+\|\tilde{L}\tilde{x}\| \|\tilde{L}\| \|\tilde{e}\|
\end{aligned}$$
In order to guarantee $\dot{V}\leq0$,enforcing $e$ to satisfy
\begin{equation*}
\|\tilde{e}\| \leq \sigma \frac{\|\tilde{L}\tilde{x}\|}{\|\tilde{L}\|}
\end{equation*}
where $\sigma\in(0,1)$. Then we have
\begin{equation*}
\dot{V}\leq (\sigma-1)\|\tilde{L}x\|^2
\end{equation*}
It is negative for $\|\tilde{L}x\|\neq0$.\\
Hence the event-trigger condition can be taken as
\begin{equation}\label{trig}
\|\tilde{e}\|=\sigma \frac{\|\tilde{L}x\|}{\|\tilde{L}\|}.
\end{equation}

\begin{theo}\label{thm1}
Consider a multi-agent system $\dot{x}=u$ with the event-based interaction law (\ref{cont}). Assume the undirected graph $\mathcal{G}$ is connected and structurally balanced. Then all agents achieve average bipartite consensus, i.e.,
$$\lim_{t\to \infty}x(t)=\frac{1}{n}\mathbf{1}^TDx(0)D\mathbf{1}.$$
\end{theo}

Proof: On the one hand, since $\dot{V}\leq (\sigma-1)\|\tilde{L}\tilde{x}\|^2$, we have  $\lim\limits_{t\to \infty}\tilde{L}\tilde{x}(t)=0$. On the other hand, since $\mathcal{G}$ is connected and structurally balanced, $\tilde{x}$ will approach to $\frac{1}{n}\mathbf{1}^T\tilde{x}(0)\mathbf{1}$ as time goes to infinity. Then it is not difficult to follow that $\lim\limits_{t\to \infty}x(t)=\frac{1}{n}\mathbf{1}^TDx(0)D\mathbf{1}$ and, furthermore, $\lim\limits_{t\to \infty}x_i(t)=\frac{1}{n}|\mathbf{1}^TDx(0)|$. The proof is thus completed. \hfill\rule{6pt}{6pt}

The event-based update schedule of the proposed interaction protocol (\ref{cont}) has a positive lower bound on the inter-event times, which is stated as below without proof:
\begin{theo}
Consider a multi-agent system $\dot{x}=u$ with the event-based interaction law (\ref{cont}). If the graph $\mathcal{G}$ is connected and structurally balanced. Then the inter-event times $\{t_{l+1}-t_l\}$ are lower bounded by
\begin{equation}
\tau=\frac{\sigma}{\|\tilde{L}\|(1+\sigma)}.
\end{equation}
\end{theo}

\subsection{Directed Signed Networks}

\begin{lemm}\label{lem2}
If a directed signed graph $\mathcal{G}$  is strongly connected and structurally balanced, all eigenvalues of the signed Laplacian matrix $L$ have nonnegative real-parts and $0$ is a simple eigenvalue.
\end{lemm}

Take a analogous variable change $\tilde{x}=Dx$ and $\tilde{e}=De$. Then one has the following closed-loop multi-agent system:
\begin{equation}\label{compe2}
\dot{\tilde{x}}(t)=-\tilde{L}\tilde{x}(t)-\tilde{L}\tilde{e}(t)
\end{equation}
Let $y=\tilde{L}x$. One has $\dot{y}=-\tilde{L}(y+\tilde{L}\tilde{e})$.

A candidate ISS Lyapunov function is given as
\begin{equation}
V(y)=\frac{1}{2}y^Ty.
\end{equation}
The derivative of $V(y)$ is
$$
\dot{V}(y)=-y^T\tilde{L}y-y^T\tilde{L}^2\tilde{e}.
$$
Define $\tilde{L}^s=\frac{1}{2}\Big(\tilde{L}+\tilde{L}^T\Big)$. Then,
$$
\begin{aligned}
\dot{V}(y)&=-y^T\tilde{L}^sy-y^T\tilde{L}^2\tilde{e}\\
&\leq -\lambda_2(\tilde{L}^s)\|y\|^2+\|y\|\|\tilde{L}\|\|\tilde{L}\tilde{e}\|.
\end{aligned}
$$

In order to guarantee $\dot{V}\leq 0$, enforcing $\tilde{L}\tilde{e}$ to satisfy
\begin{equation}
\|\tilde{L}\tilde{e}\| \leq \sigma\lambda_2(\tilde{L}^s)\frac{\|y\|}{\|\tilde{L}\|}
\end{equation}
where $\sigma\in(0,1)$.
Then we have
\begin{equation*}
\dot{V}\leq (\sigma-1)\lambda_2(\tilde{L}^s)\|y\|^2.
\end{equation*}
It is negative for $y=\tilde{L}x\neq 0$. Hence the event-trigger function is:
\begin{equation}\label{trig1}
\|\tilde{L}e\|=\sigma\lambda_2(\tilde{L}^s)\frac{\|\tilde{L}x\|}{\|\tilde{L}\|}.
\end{equation}

Then another main result follows.
\begin{theo}\label{thm2}
Consider a multi-agent system $\dot{x}=u$ with the event-based interaction law (\ref{cont}). Assume the directed graph $\mathcal{G}$ is strongly connected and structurally balanced. Then all the agents achieve a bipartite consensus,i.e.,
$\lim_{t\to \infty}x(t)=\alpha^TD x(0) D\mathbf{1}$, where $\alpha^TDL=0$ and $\alpha^T\mathbf{1}=1$. At the same time, the inter-event times $\{t_{l+1}-t_l\}$ are lower bounded by
\begin{equation}
\tau=\frac{\sigma\lambda_2(\tilde{L}^s)}{\|\tilde{L}\|(\|\tilde{L}\|+\sigma)\lambda_2(\tilde{L}^s)}.
\end{equation}
\end{theo}

Proof: The proof is similar with that in Theorem \ref{thm1} and omitted here. \hfill\rule{6pt}{6pt}

\section{Simulation Results}\label{simu}

Some simulations are given to illustrate the collective event-based dynamics of multi-agent systems over a signed network.

\begin{exam}
Consider an undirected antagonistic network with ten agents whose Laplacian matrix is given by
\begin{equation*}
\mathbf{L} =
\left( \begin{array}{cccccccccc}
3 &	-1&	0&	0&	0&	-1&	1&	0&	0&	0\\
-1&	2&	-1&	0&	0&	0&	0&	0&	0&	0\\
0&	-1&	2&	-1&	0&	0&	0&	0&	0&	0\\
0&	0&	-1&	3&	-1&	0&	0&	1&	0&	0\\
0&	0&	0&	-1&	4&	-1&	1&	1&	0&	0\\
-1&	0&	0&	0&	-1&	4&	1&	1&	0&	0\\
1&	0&	0&	0&	1&	1&	5&	-1&	0&	-1\\
0&	0&	0&	1&	1&	1&	-1&	5&	-1&	0\\
0&	0&	0&	0&	0&	0&	0&	-1&	2&	-1\\
0&	0&	0&	0&	0&	0&	-1&	0&	-1&	2
\end{array} \right)
\end{equation*}
Obviously,the agents $v_1,\dots,v_6$ and $v_7,\dots,v_{10}$ belong to two adversary groups. Take $\sigma=0.9$. Ten agents start form random initial conditions and evolve under the control law (\ref{cont}).

Figure~\ref{un:tr} shows the evolution of the error norm in the case of undirected networks.The blue line represents the evolution of the error norm $\|e(t)\|$, which stays below the threshold $\|e\|_{max}=\sigma\frac{\tilde{L}x}{\tilde{L}}$ which is represented by the green line.The existence of a minimum inter-event time is visible in this example. Figure~\ref{un:st} shows the evolution of the agents' state in the case of undirected networks. The red line represents the evolution of agents $v_1,\dots,v_6$ and the blue line represents the evolution of agents $v_7,\dots,v_{10}$. All the agents finally converge to two separate final states.

\begin{figure}[!htbp]
\centering
\includegraphics[width=0.95\columnwidth]{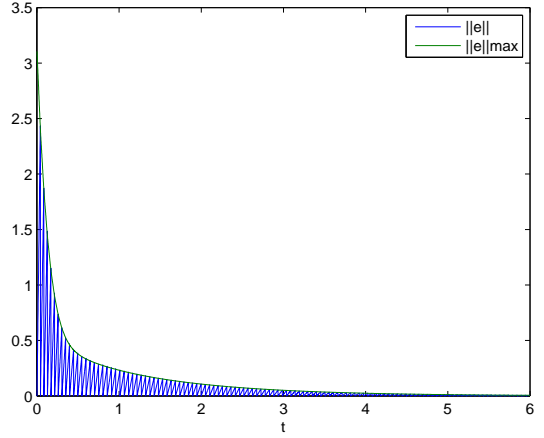}
\caption{Evolution of the error norm in the case of undirected networks}
\label{un:tr}
\end{figure}

\begin{figure}[!htbp]
\centering
\includegraphics[width=0.95\columnwidth]{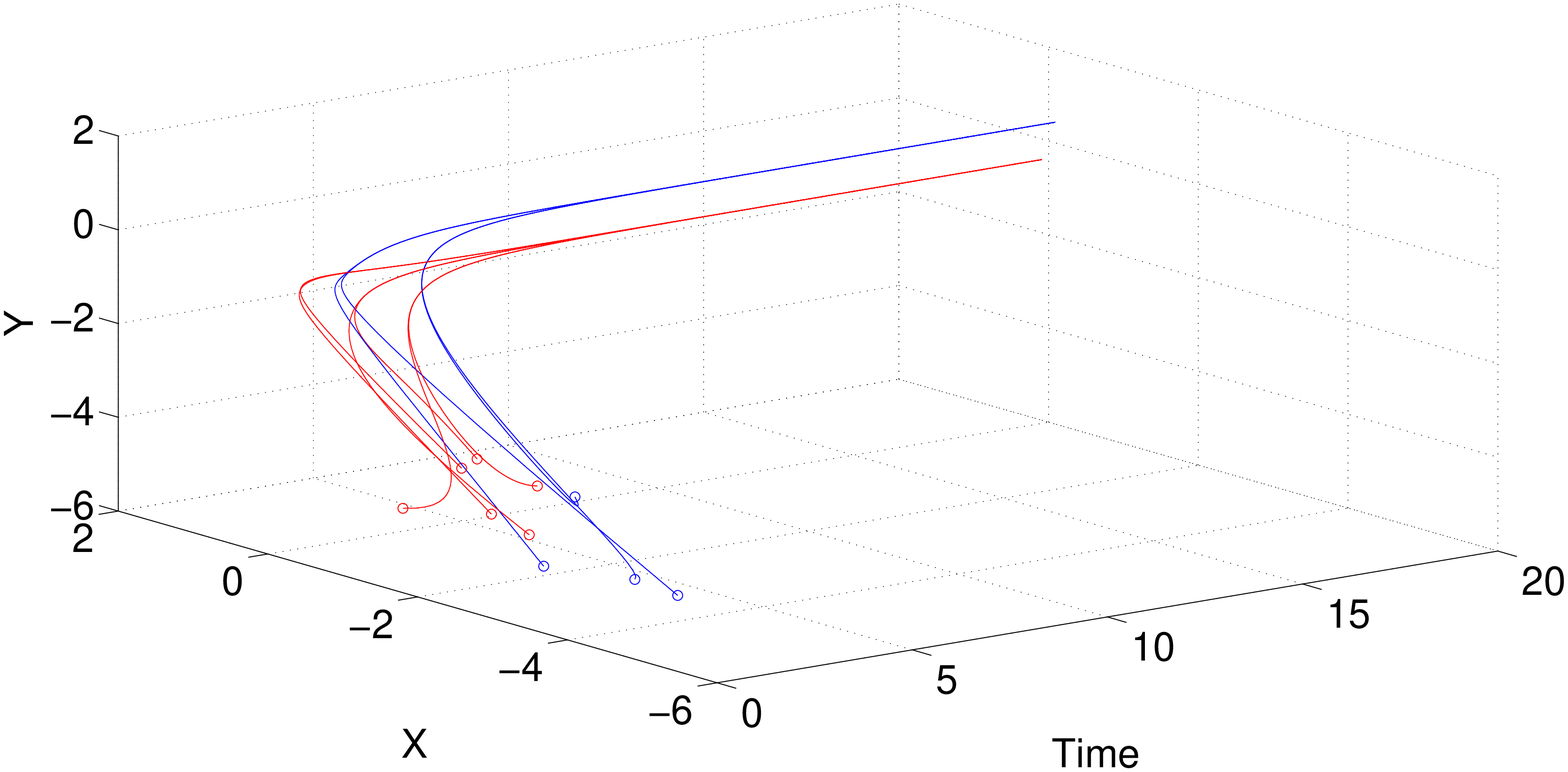}
\caption{Evolution of the collective state in the case of undirected networks}
\label{un:st}
\end{figure}

\begin{figure}[!htbp]
\centering
\includegraphics[width=0.95\columnwidth]{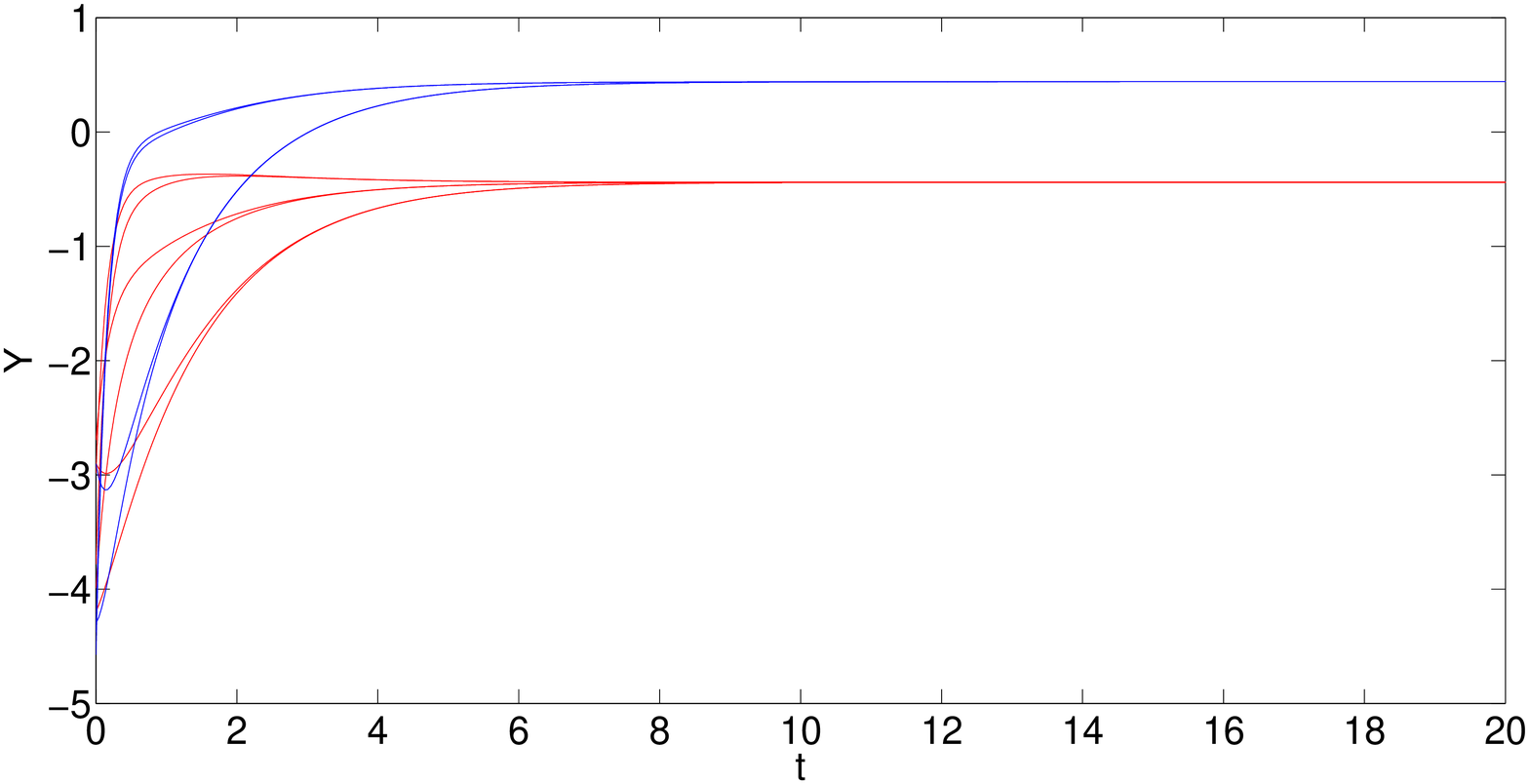}
\caption{Evolution of the y-axis state in the case of undirected networks}
\label{un:st1}
\end{figure}
\end{exam}

\begin{exam}
Consider a directed antagonistic network with ten agents whose signed Laplacian matrix is given as
\begin{equation*}
\mathbf{L} =
\left( \begin{array}{cccccccccc}
1&	0&	0&	-1&	0&	0&	0&	0&	0&	0\\
-1&	1&	0&	0&	0&	0&	0&	0&	0&	0\\
0&	-1&	2&	0&	0&	0&	0&	1&	0&	0\\
0&	0&	0&	1&	-1&	0&	0&	0&	0&	0\\
0&	0&	0&	0&	1&	-1&	0&	0&	0&	0\\
0&	0&	-1&	0&	0&	1&	0&	0&	0&	0\\
0&	0&	1&	0&	0&	0&	2&	0&	-1&	0\\
0&	0&	0&	0&	0&	0&	-1&	2&	-1&	0\\
0&	0&	0&	0&	0&	0&	-1&	0&	2&	-1\\
0&	0&	0&	0&	0&	0&	0&	-1&	0&	1
\end{array} \right)
\end{equation*}
Obviously,the agents $v_1,\cdots,v_6$ and $v_7,\cdots,v_{10}$ belong to two adversary groups. The interaction network $\mathcal{G}$ is strongly connected and structurally balanced. Set $\sigma=0.9$. Ten agents start form random initial conditions and evolve under the interaction law (\ref{cont}).

Figure~\ref{dtr} shows the evolution of the error norm in the case of directed networks. The blue line represents the evolution of the error norm $\|\tilde{L}e\|$, which stays below the threshold $\|\tilde{L}e\|_{max}=\sigma\lambda_2(\tilde{L}^s)\frac{\tilde{L}x}{\tilde{L}}$. The bound is represented by the green line. The existence of a minimum inter-event time is visible in this example. Figure~\ref{dst} shows the evolution of the collective state in the case of directed networks. The red line represents the evolution of agents $v_1, \cdots, v_6$ and the blue line represents the evolution of agents $v_7, \cdots, v_{10}$.From the Figure,the nodes finally converge to two separate states.

\begin{figure}
\begin{center}
\includegraphics[width=0.95\columnwidth]{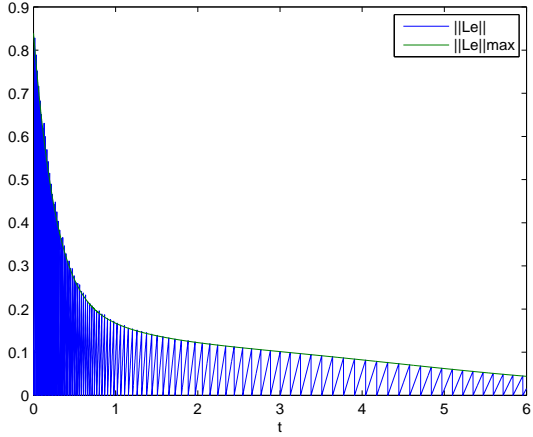}
\end{center}
\caption{Evolution of the error norm in the case of directed networks}
\label{dtr}
\end{figure}

\begin{figure}
\begin{center}
\includegraphics[width=0.95\columnwidth]{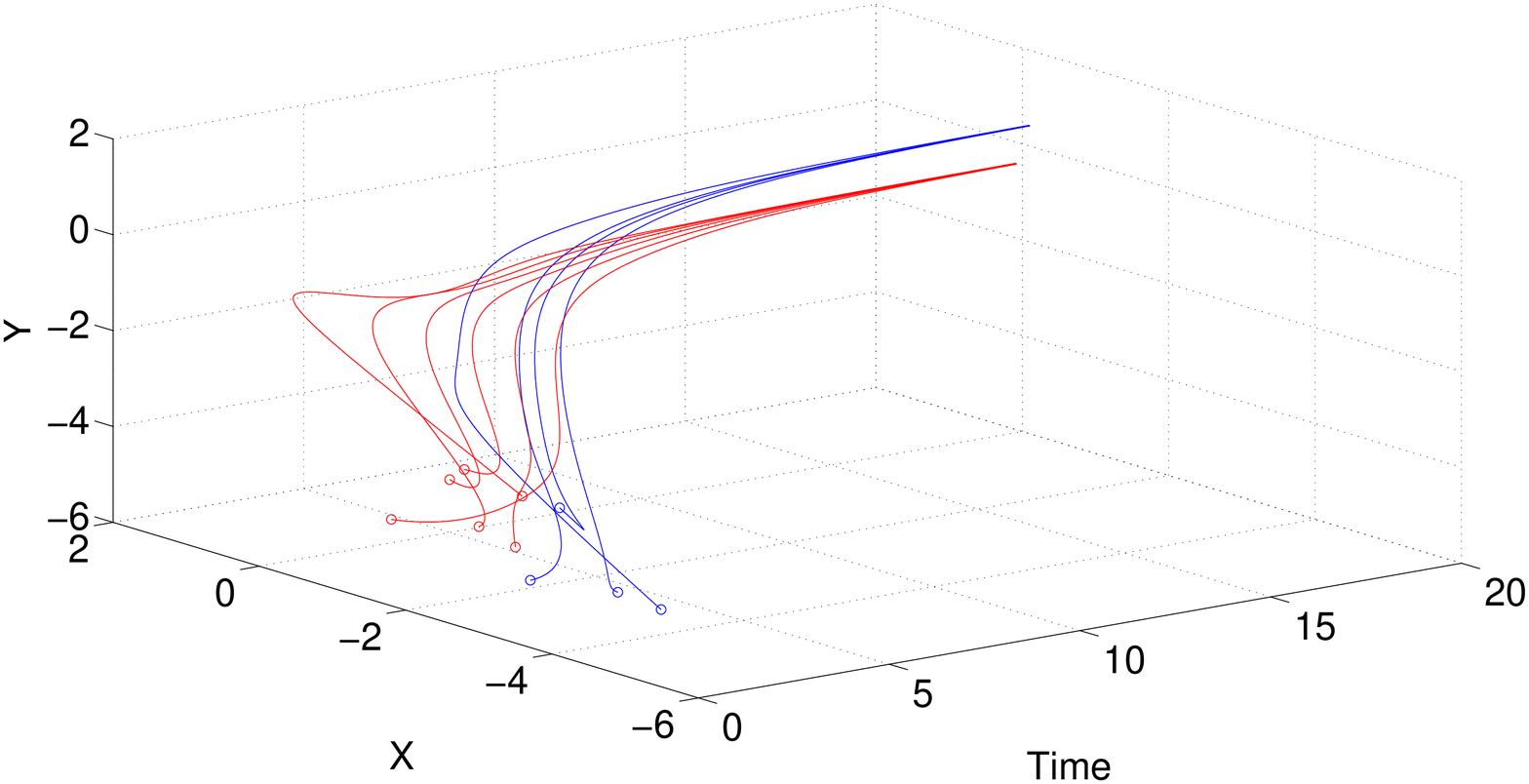}
\end{center}
\caption{Evolution of the collective state in the case of directed networks}
\label{dst}
\end{figure}

\begin{figure}[!htbp]
\centering
\includegraphics[width=0.95\columnwidth]{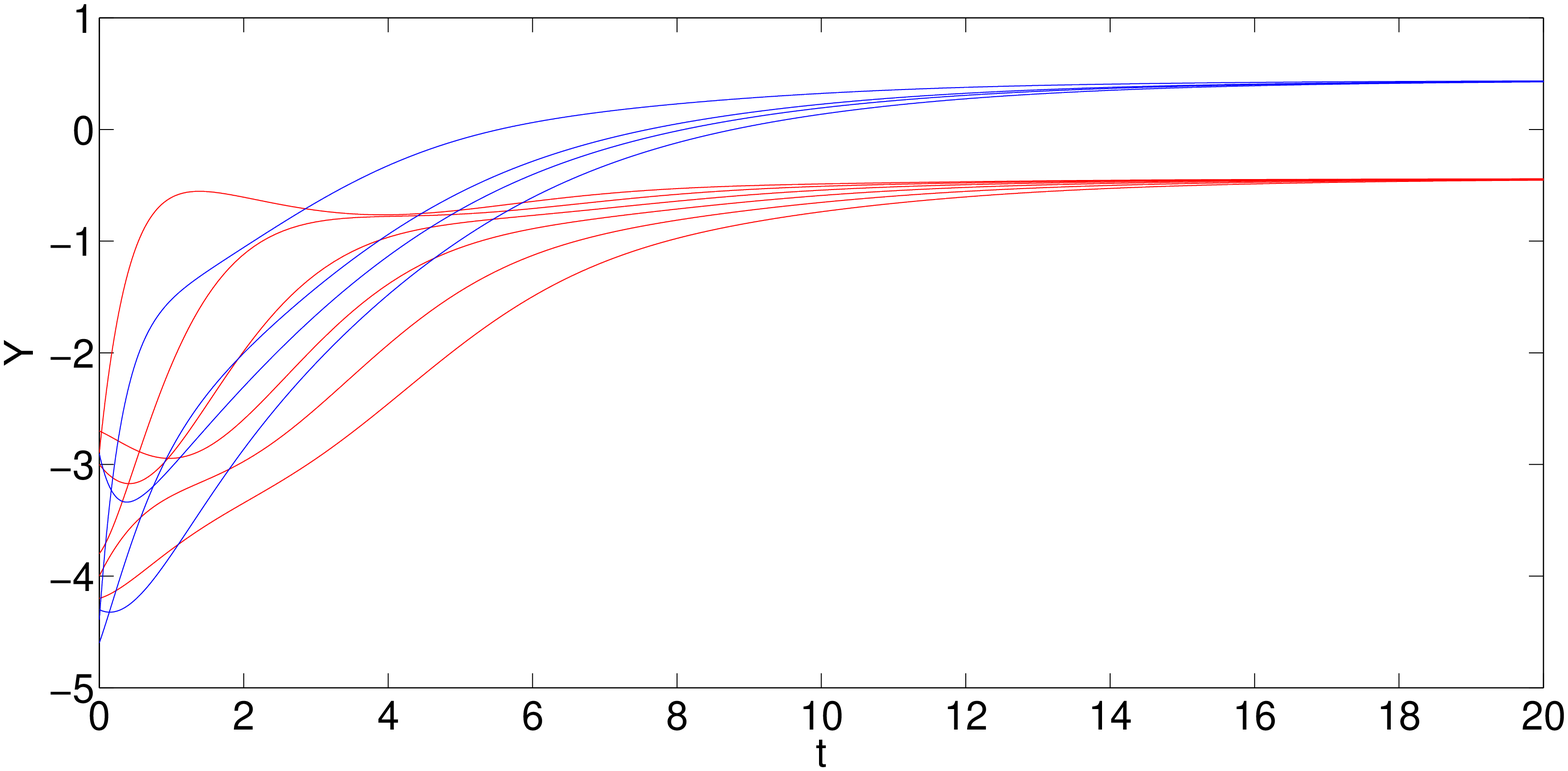}
\caption{Evolution of the y-axis state in the case of undirected networks}
\label{dst1}
\end{figure}
\end{exam}

\section{Conclusion}\label{conc}

In this paper,a new event-based consensus control problem has been investigated for multi-agent systems with antagonistic interactions. A bipartite consensus problem was solved by using an event-based update strategy for agents with cooperative and adversary relationships. When the multi-agent network is undirected or directed, some structurally balanced conditions have been given to ensure all the agents to achieve a bipartite consensus. At the same time, the evolution of the measurement error has also been illustrated to validate the event-based consensus protocols. Future interests include bipartite consensus over time-varying multi-agent networks.

\section*{Acknowledgment}

This work is supported by the National Natural Science Foundation of China under Grant 61104104 and the Scientific Research Foundation for the Returned Overseas Chinese Scholars, State Education Ministry of China.

%\balance

\end{document}